\documentclass[showpacs, amsmath, amssymb, superscriptaddress,
 pra, twocolumn, notitlepage]{revtex4-1}
\usepackage{graphicx}
\def\({\left(}
\def\){\right)}
\def\[{\left[}
\def\]{\right]}

\def\vector#1{\mbox{\boldmath $#1$}}

\newcommand{\braket}[2]{\left\langle #1 | #2\right\rangle}

\newcommand{\bra}[1]{\left\langle #1 \right|}
\newcommand{\ket}[1]{\left| #1 \right\rangle}
\newcommand{\tr}{\mathrm{tr}}

\begin{document}

\title{Secret key rate of a continuous-variable quantum-key-distribution scheme when  the detection process is inaccessible to eavesdroppers}

\author{Ryo Namiki}
\affiliation{Department of Physics, Gakushuin University, 1-5-1 Mejiro, Toshima-ku,  Tokyo 171-8588, Japan}
\author{Akira Kitagawa}
\affiliation{Faculty of Education, Kochi University, 2-5-1 Akebono-cho, Kochi 780-8520, Japan}
\author{Takuya Hirano}
\affiliation{Department of Physics, Gakushuin University, 1-5-1 Mejiro, Toshima-ku,  Tokyo 171-8588, Japan}


\begin{abstract}
We have developed a method to calculate a secret key rate of a continuous-variable quantum key distribution  scheme using four coherent states and postselection for a general model of Gaussian attacks. We assume that the transmission line and detection process are described by a pair of Gaussian channels. 
In our analysis, while the loss and noise on the transmission line are induced by an eavesdropper, Eve,  who can replace the transmission line with a lossless and noiseless optical fiber, she is assumed  inaccessible to the detection process.  
By separating the transmission noise and detection noise, we can always extract a larger key compared with the case that all loss and noises are induced by eavesdropper's interference. An asymptotic key rate against collective Gaussian attacks can be determined numerically for given channels' parameters. The improvement of the key rates turns out to be more significant for the reverse-reconciliation scheme. 
\end{abstract}
 

\maketitle


\section{Introduction}
Quantum key distribution (QKD) protocols enable two remote parties, usually called Alice and Bob, to share a random bit sequence called the secret key  \cite{rmp-QKD09XX,Lo2014,diam16}. 
 Based on the laws of physics,  the secret key can be proven information theoretically secure from an eavesdropper, Eve,  who can access the transmission channels between Alice and Bob with possibly unlimited technology and computational power.  There have been wide activities to demonstrate QKD beyond the proof-of-principle experiments, and several field tests have been reported \cite{Sasaki11,SwissQKD12,Shimizu14,Wang14,Tang2015,Dixon:15,Dixon2017,Sasaki17,Bunandar18,Tajima17,CVFieldTestNJPSVQKD09,CVHuang:16}.

In contrast to several  no-go theorems in Gaussian continuous-variable (CV) quantum information protocols \cite{Eisert2002,Fiurasek,Giedke2002,Bra05}, CV-QKD protocols enjoy quantum optical homodyne measurements and the properties of coherent states to offer possible solution in quantum safety \cite{DiamantiEntropy}. In addition to 
 the study 
of quantum optical methods, CV-QKD schemes potentially have advantage in practical implementation. They could be operated with commercially available detectors and deployed in lit fiber networks possibly with the presence of classical channels \cite{1367-2630-12-10-103042,Kumar15,Nakazawa17}.  An important theoretical challenge is to establish mathematical proof of general safety \cite{Lev17}. It would be crucial to address how to generate secure local oscillator signals  \cite{PhysRevX.5.041009,PhysRevX.5.041010,Huang:15,Laud17,PhysRevA.95.012316}.

Physically, an essential point in QKD protocols is that quantum mechanical states are employed so that Eve's effort to read the transmission signal unavoidably induces errors on Bob's observation of the transmitted states.  The amount of the secret key is usually calculated conservatively assuming that any observed signal loss and noises are induced by Eve. 
 In practice, a portion of  the loss and noises is induced at Bob's detector. Since the detector is inside Bob's station it is feasible to assume that  the detection process is inaccessible to Eve \cite{rmp-QKD09XX}. Therefore, if one can know how much part of the decoherence is given at the detector, a better key rate will be estimated by using the fact that Eve cannot extract correlation from the noise added inside Bob's station. Such scenario has been investigated in discrete-variable QKD schemes \cite{Boil05,Hayashi07}. In CV-QKD schemes, Gaussian description of state-evolution is a basic approach to model the system performance. In the Gaussian modulated CV-QKD scheme,  Eve's inaccessibility to the detector noise has been readily taken into account assuming the action of Gaussian channels \cite{Lodewyck07,CVFieldTestNJPSVQKD09,Jougue11,Jouguet12,Us16}. However, these effects have little  been studied in types of CV-QKD schemes which use discrete modulation and postselection \cite{Sil02,Lor04,Hirano2003,heid06a,Namiki2006,Ichikawa17}. 

For the postselection protocols of Refs.~\cite{Sil02,heid06a,Hirano2003}, an asymptotic key rate against collective Gaussian attacks was determined both in the direct reconciliation (DR) and  reverse reconciliation (RR) scenarios \cite{Heid07,Hirano17}. For these discrete-modulation CV-QKD protocols, it has been unknown that a Gaussian attack could be the optimal attack, and the key rate against general attacks could be much lower than the key rate against Gaussian attacks \cite{Zhao09}.  Note that another type of discrete-modulation CV-QKD protocols which does not employ postselection has been investigated in Refs.~\cite{PhysRevLett.102.180504,PhysRevLett.106.259902,PhysRevA.83.042312}.

In this paper, we investigate the security of the four state CV-QKD protocol \cite{Hirano2003}
 when the loss and noises in the detection process are inaccessible to Eve. 
We derive a formula to determine Eve's density operator when the system describing detector's loss and noises is simply traced out. Using this density operator the key rate against collective Gaussian attacks can be calculated.

This paper is  organized as follows. In Sec.~\ref{pre}, we introduce basic notions on our protocol.  In Sec.~\ref{eca}, we describe the model of Eve's attack  and how to determine her density operator. 
In Sec.~\ref{keyrates} we show numerically calculated key rates for a couple of noise parameters.   
We conclude this paper in Sec.~\ref{Conclusion}.

\section{Protocol and notation}
\label{pre}

We consider the four coherent-state protocol \cite{Hirano2003}. We basically use the same notation as in Ref.~\cite{Hirano17}.  Alice randomly sends one of four coherent states $\ket{S}\in  \{ \ket{\pm\alpha}, \ket{\pm i\alpha}\} $ with $\alpha>0$ to Bob. Bob performs quadrature measurement either $x$-basis 
or $p$-basis: 
\begin{align}
x=\frac{a+a^\dag}{2},
\qquad
p=\frac{a-a^\dag}{2i}, 
\end{align} where the commutation  relation $[ a, a^\dag] =1 $ is assumed to hold. For notation convention, we may refer to the optical mode transmitted from Alice to Bob as  the mode $B$.  
From the sequence that Alice sent $\ket{\pm \alpha }$ and Bob measured the $x $ quadrature, sifted key is generated by assigning bit value ``0'' for Bob's measurement outcome $m>0$ and bit value ``1'' for   $m<0$. The same procedure is executed for the sequence that Alice sent $\ket{ \pm i\alpha }$ and Bob measured the $p $  quadrature. Bob may announce the absolute values of his measurement outcomes $|m|$, and further classical key agreement procedure will be carried out by using the index $|m|$. 
Due to the phase-space symmetry with $\pi/2$ rotation, it is sufficient to consider the case that Alice sends $\ket{S} = \ket{ \pm \alpha } $ and Bob obtains an outcome $m$ associated with his measurement of the $x$ quadrature.  We assume linear loss and symmetric Gaussian excess noise through the transmission and detection of the signal. This implies that Bob's quadrature distribution when Alice's signal is $S$ can be written as 
\begin{align}
P(m|S)=\sqrt{\frac{2}{\pi(1+\xi)}}e^{-2\frac{(m-\sqrt{\eta}S)^2}{1+\xi}},
\label{ms}
\end{align}
where $\eta $ is the total transmission and $\xi$ is the total excess noise.  Note that this distribution reduces to the quadrature distribution of the coherent state $\ket{\alpha }$ when there is no loss and no excess noise, namely, we have $P(x|\alpha)=|\braket{x}{\alpha}|^2$ when $(\eta,\xi)=(1,0)$.

Conditioned on the absolute value of Bob's outcome $|m|$, the transmission is considered to  be a binary symmetric channel with the bit error rate 
\begin{align}
\epsilon_{|m|}&:= \frac{P(-|m||\alpha)}{P(m|\alpha)+P(-m|\alpha)}\nonumber\\
&= \frac{P(|m||-\alpha)}{P(m|-\alpha)+P(-m|-\alpha)}\nonumber\\
&= \[1+e^{8\frac{\sqrt{\eta}}{1+\xi}|m|\alpha}\]^{-1}. \label{BERep}
\end{align}
From the bit error rate $\epsilon$,  the mutual information between Alice and Bob can be calculated as 
\begin{align}
I_{\rm AB}=1-f h(\epsilon). \label{MIAB}
\end{align}
where $h(\epsilon)=-\epsilon\log_2\epsilon-(1-\epsilon)\log_2(1-\epsilon)$ is the binary entropy function 
and $f\ge 1$ represents 
  an efficiency of error correction.  In what follows, we assume an ideal error correction and set $f=1$.

\section{Interaction Model and Eve's knowledge} \label{eca}

\begin{figure}[t]
   \centering
 \includegraphics[width=\linewidth]{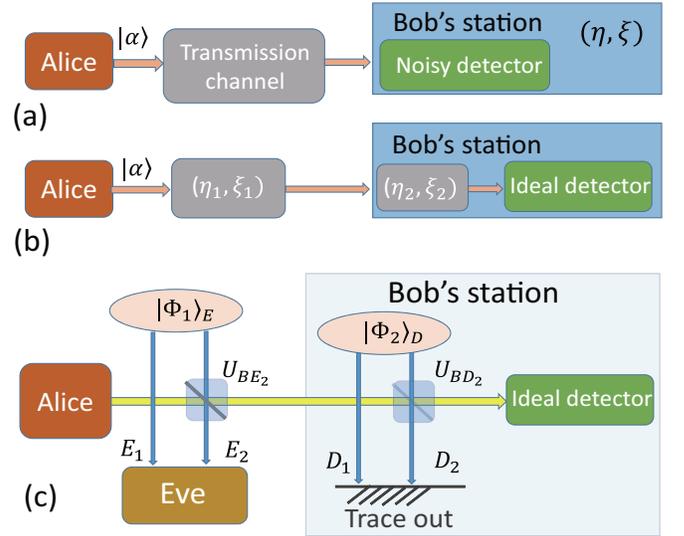} \\
\caption{(Color online)(a) Alice sends a  coherent state $\ket{\alpha}$ through a lossy and noisy channel. Bob observes quadrature with the total transmission $\eta $ and the total excess noise $\xi$. (b) The transmission channel is modeled by a lossy and noisy Gaussian channel with the transmission $\eta_1$ and the excess noise $\xi_1$. Bob's detector is  modeled by another lossy and noisy Gaussian channel with the transmission $\eta_2$ and the excess noise $\xi_2$ followed by an ideal homodyne detector. (c) The action of Gaussian channels $(\eta_i,\xi_i)$ with $i=1,2$ can be described by beamspliter unitaries $U_{BE_2}$ and $U_{BD_2}$  coupling to two-mode squeezed states, $\ket{\Phi_1}_E$ and $\ket{\Phi_2}_D$.}
   \label{modelfig}
 \end{figure}
We consider the physical model of the signal transmission in Fig.~\ref{modelfig}. 
Let us assume the signal is transmitted from Alice to Bob through a lossy and noisy Gaussian channel with the transmission $\eta_1$ and excess noise $\xi_1$. The channel action is virtually described  as follows: The signal is mixed with an ancillary mode $E_2$ of a two-mode squeezed state $\ket{\Phi_1}_{E_1E_2}$ at a beamsplitter. Then, the extra modes ${E_1E_2}$ are traced out. We assume that 
Eve can access  any signal information leaking out at the transmission line. This condition can be described by assuming that Eve holds the extra modes  ${E_1E_2}$. The interaction model with a two-mode squeezed state and a beamsplitter is often called the \textit{entangling cloner} \cite{Gross03}.  Due to the properties of the two-mode squeezed state, an entangling cloner induces linear loss and symmetric Gaussian noise on the signal mode in such a way that Eve holds correlation on the signal noise. There is another interaction model to simulate this effect with a two-mode squeezer and two beamsplitters \cite{Namiki2006a}.

Let us also assume the signal is transmitted through another Gaussian channel with the transmission $\eta_2$ and excess noise $\xi_2$ at Bob's detection process in Bob's station.  We assume that Eve is unable to access  any signal information leaking out at the detection process inside Bob's station. The signal is mixed with an ancillary mode $D_2$ of a two-mode squeezed state $\ket{\Phi_2}_{D_1D_2}$ at a beamsplitter. Then, the extra modes ${D_1D_2}$ are traced out, and Eve has no chance to hold the purifying modes  ${D_1D_2}$.  
To this end,  our model contains two entangling cloners associated with the pair of Gaussian channels. One is used to describe Eve's access. The other one associated with the extra modes $D_1D_2$ is used to describe a realistic detection process .

Let us write Eve's initial two-mode squeezed state as 
\begin{align}
\ket{\Phi_1}_E&&=\sqrt{\frac{2}{\pi}}\int_{-\infty}^\infty\int_{-\infty}^\infty dx_1dx_2\ e^{-V_1 x_1^2-x_2^2/V_1}\nonumber\\
&&\times\ket{\frac{x_1+x_2}{\sqrt{2}}}_{E_1}\ket{\frac{x_1-x_2}{\sqrt{2}}}_{E_2}
\end{align}
where  $\ket{x}_{E_i}$ is the eigenket 
of a quadrature operator $x$ of the mode $E_i$ with eigenvalue $x$.
The parameter  $V_1$ is  associated with the channel's  parameters as   \cite{Symul07}
\begin{align}\frac{1}{2}\(V_	1+\frac{1}{V}_1\)=\frac{1-\eta_1+\xi_1}{1-\eta_1}\label{V1}.
\end{align}Since the coherent state $\ket{S}$ can be written as
\begin{align}
\ket{S}=\(\frac{2}{\pi}\)^{\frac{1}{4}}\int_{-\infty}^{\infty}dx\, e^{-(x-S)^2}\ket{x}
\end{align}
and the beam splitter of transmission $\eta$ leads to the transformation
\begin{align}
\ket{x}|x_2\rangle_{E_2}\to|\sqrt{\eta}x-\sqrt{1-\eta}x_2\rangle|\sqrt{1-\eta}x+\sqrt{\eta}x_2\rangle_{E_2}, \label{bst1}
\end{align}the state after the beam-spliter interaction $U^{\rm (BS)}$  takes the form of  \cite{Hirano17} 
\begin{align}U_{\rm BE2}^{\rm (BS)}\ket{S}_B\ket{\Phi_1}_E= \int_{-\infty}^\infty dm\ket{m}_B\ket{\psi(S, m)}_E,
\end{align}
where
\begin{align}\ket{\psi(S,m)}=&\(\frac{8}{\pi^3\eta_1^2}\)^{\frac{1}{4}}\int_{-\infty}^\infty\int_{-\infty}^\infty dx_1dx_2\, \psi(S,m)\nonumber\\
& \times \ket{\frac{x_1+x_2}{\sqrt{2}}}_{E_1}\ket{\sqrt{\frac{1-\eta_1}{\eta_1}}m+\frac{x_1-x_2}{\sqrt{2\eta_1}}}_{E_2}
\end{align}
with
\begin{align}\psi(S,m)=e^{-\[\sqrt{\frac{1-\eta_1}{2\eta_1}}\(x_1-x_2\)+\frac{m}{\sqrt{\eta_1}}-S\]^2-V_1x_1^2-x_2^2/V_1}.
\end{align}

Let us write the initial state of detector's environment for the second entangling cloner as  
\begin{align}
\ket{\Phi_2}_D&&=\sqrt{\frac{2}{\pi}}\int_{-\infty}^\infty\int_{-\infty}^\infty dy_1dy_2\ e^{-V_2 y_1^2-y_2^2/V_2}\nonumber\\
&&\times\ket{\frac{y_1+y_2}{\sqrt{2}}}_{D_1}\ket{\frac{y_1-y_2}{\sqrt{2}}}_{D_2}
\end{align}
where the parameter  $V_2$ is associated with detector's parameters as  
\begin{align}\frac{1}{2}\(V_2+\frac{1}{V_2}\)=\frac{1-\eta_2+\xi_2}{1-\eta_2} \label{V2}
\end{align}
Using the transformation Eq.~\eqref{bst1}, we can write the state before the ideal quadrature measurement device as 
\begin{align}
&U_{\rm BD2}^{\rm (BS)} U_{\rm BE2}^{\rm (BS)} \ket{S}_B\ket{\Phi_1}_E\ket{\Phi_2}_D=\int_{-\infty}^\infty dm\ket{m}_B\ket{\psi^\prime (S, m)}_{ED}, \label{unitarysan} \end{align}
where 
\begin{align}
& \ket{\psi^\prime (S, m)}  \nonumber \\ 
=& \sqrt{\frac{2}{\pi}}\int_{-\infty}^\infty\int_{-\infty}^\infty\frac{ dY_1 dY_2 }{\sqrt{\eta_2}}\ket{\psi(S, \frac{m}{\sqrt{\eta_2}}+\sqrt\frac{{1-\eta_2}}{ \eta_2 }Y_2)}_E  \ket{Y_1}_{D_1}  \nonumber \\
&\otimes \ket{\sqrt{\frac{1-\eta_2}{\eta_2}}m+\frac{Y_2}{\sqrt{ \eta_2}}}_{D_2} e^{-\frac{V_2}{2}(Y_1+Y_2)^2 -\frac{(Y_1-Y_2)^2}{2V_2}}. \end{align}
Since Eve cannot access the information leakage coming from detector's environment system $D$, 
we trace out the system $D$ in what follows.

Supposing  Bob's homodyne outcome is $m$, we obtain  Eve's density operator: \begin{align}
\rho_{S, m }:=&\tr_{D_1D_2} \ket{\psi^\prime (S, m)}\bra{\psi^\prime (S, m)} 
\nonumber \\ 
=&
\sqrt{\frac{4}{\pi} \frac{1}{V_2+\frac{1}{V_2}} } \int dY  \ket{\varphi _{S,m} }\bra{\varphi _{S,m} } e^{-\frac{4 Y^2}{V_2+\frac{1}{V_2}}} \label{RhoSm}\end{align}
where 
\begin{align}
\ket{\varphi _{S,m} }= &\left(\frac{8}{\pi^3 \eta_1^2 \eta_2} \right) ^{1/4} \iint  dX_1 dX_2 
\nonumber \\ & \times e^{-\[\sqrt{\frac{1-\eta_1}{ \eta_1}}X_2 +\frac{1}{\sqrt{\eta_1 \eta_2}}(m + \sqrt{1- \eta_2} Y)-S\]^2}  
\nonumber \\ & \times e^{-\frac{V_1}{2}(X_1+X_2 )^2 -\frac{(X_1-X_2 )^2}{2V_1}  }\ket{X_1}_{E_1} \nonumber \\
& \otimes \ket{\frac{X_2}{\sqrt{\eta_1}} +\frac{\sqrt{1- \eta_1}}{\sqrt{\eta_1 \eta_2}}(m+ \sqrt{1-\eta_2} Y ) }_{E_2} \nonumber \\
= & 
 \iint  dx_1 dx_2 
\varphi_Y (x_1,x_2)\ket{x_1}_{E_1} \ket{ {x_2} }_{E_2}.  \label{WF1}\end{align}
Here, in the final expression,  we defined 
\begin{align}
\varphi_Y (x_1,x_2) := &\left(\frac{8}{\pi^3 \eta_2} \right) ^{\frac{1}{4}}  
e^{-\[\sqrt{ {1-\eta_1} } x_2 +\sqrt{\frac{\eta_1}{{ \eta_2}}}( m +  \sqrt{1- \eta_2} Y)-S\]^2}  &
\nonumber \\ & \times e^{-\frac{V_1}{2} \left[x_1+\sqrt{\eta_1}  x_2 -\sqrt\frac{1- \eta_1}{\eta_2} (m+ \sqrt{1-\eta_2} Y ) \right]^2  }&
\nonumber \\ & \times e^{-\frac{1}{2V_1} \left[  x_1-\sqrt{\eta_1} x_2  + \sqrt  \frac{1- \eta_1}{\eta_2} (m+ \sqrt{1-\eta_2} Y ) \right]^2    }. \end{align}
Further tracing out $\rho_{S,m} $ in Eq.~\eqref{RhoSm} we obtain the probability that Bob gets $m$ conditioned on Alice sends $\ket{S}$: 
\begin{align}
\tr \rho_{S,m}= \sqrt{ \frac{2}{\pi}}\sqrt{ \frac{1}{1+ \xi_1 \eta_2 +\xi_2 }}e^{-2 \frac{(m- \sqrt{\eta_1 \eta_2} S)^2}{1+ \xi_1 \eta_2 +\xi_2  } }.
\end{align} Comparing this relation with Eq.~\eqref{ms} we find
\begin{align}
\eta=& \eta_1 \eta_2, \nonumber \\ 
 \xi=  &\xi_1 \eta_2 +\xi_2. \label{CPconcate}
\end{align} This is the central relation to associate  the total channel parameters $(\eta, \xi)$ with the parameters of the transmission line $(\eta_1,\xi_1)$ and the parameters of the detection process in Fig.~\ref{modelfig}a  and Fig.~\ref{modelfig}b.  
From  the expression of $\rho_{S,m} $ in Eq.~\eqref{RhoSm}, we define normalized states:
\begin{align}
\omega_{i,j}=  \frac{ \rho_{ (-1)^i |S|,(-1)^j|m|}}{\tr \rho_{ (-1)^i |S|,(-1)^j|m|}}, \label{unnorij}
\end{align} where $i,j \in \{0,1\}$ can be associated with Alice's bit and Bob's bit, respectively . 
As we can see from Eq.~\eqref{RhoSm},  Eve's state $\rho_{S,m}$ is susceptible to both Alice's state preparation $S$ and Bob's measurement outcome $m$. In fact, Eve's  information can be bounded by the difference of her states due to the change of the signs of $S$ and $m$, and the normalized state $\omega_{i,j}$ in Eq.~\eqref{unnorij} helps us to express the relevant information quantities  \cite{Heid07,Hirano17}.

Let us recall the bit error rate $\epsilon $ is given in Eq.~\eqref{BERep}.   
In the case of the DR scheme, Eve tries to know   Alice's preparation of bit. Eve's knowledge is determined from the difference of her states with Alice's choice $i = \{0,1\}$,
\begin{align}
 &\rho^0_{\rm A}=(1-\epsilon) \omega_{00} +\epsilon\omega_{01},\nonumber\\
& \rho^1_{\rm A}=(1-\epsilon)\omega_{11}+\epsilon\omega_{10}.
\label{dmA}
\end{align}
In the case of the RR scheme, Eve tries to know   Bob's bit. Eve's knowledge is thus determined from the difference of her states with the sign of Bob's outcome $j = \{0,1\}$,
\begin{align}
&\rho^0_{\rm B}=(1-\epsilon) \omega_{00}+\epsilon\omega_{10},\nonumber\\
&\rho^1_{\rm B}=(1-\epsilon) \omega_{11}+\epsilon \omega_{01}.
\label{dmB}
\end{align}
Therefore, the information potentially accessible to Eve is bounded from above by the Holevo quantity 
\begin{align}
\chi=
\begin{cases}
S(\rho)-S(\rho_{\rm A}^0)/2-S(\rho_{\rm A}^1)/2      & \text{for DR}, \\
S(\rho)-S(\rho_{\rm B}^0)/2-S(\rho_{\rm B}^1)/2      & \text{for RR},
\end{cases}
\label{hol}
\end{align}
where we define 
$\rho:=(\rho_{\rm A}^0+\rho_{\rm A}^1)/2=(\rho_{\rm B}^0+\rho_{\rm B}^1)/2$
and  the von Neumann entropy is given by 
$S(\rho)=-{\rm Tr}(\rho\log_2\rho) $.

Although we have proceeded our calculation based on the notation convention of Refs.~\cite{Heid07,Hirano17}, it may be instructive to remind the connection  to an entanglement based picture. In such a picture,  Alice prepares a qubit-mode entangled state $\ket{\phi}_{AB}=(\ket{0}_A\ket{\alpha}_B+\ket{1}_A\ket{-\alpha}_B)/{\sqrt 2}$, and sends the mode $B$ to Bob. Note that the whole system  consists of  Alice, Bob, Eve, and detector's environment, which are respectively denoted by the subscripts $A$, $B$, $E$, and $D$. Using the two-mode squeezed states $\ket{\Phi_{1}}$,  $\ket{\Phi_{2}}$, and the unitary operators in Eq.~\eqref{unitarysan},  the whole system can be written as a pure state on  the four modes,  
\begin{align}
\ket{\Psi}_{ABDE} =  U_{\rm BD2}^{\rm (BS)} U_{\rm BE2}^{\rm (BS)} \ket{\phi}_{AB} \ket{\Phi_1}_E\ket{\Phi_2}_D. 
\end{align}
Then,  the tripartite state between Alice, Bob, and Eve, is given by tracing out detector's environment,  
\begin{align}
\rho_{ABE} = \tr_D \left(  \ket{\Psi} \bra{\Psi} 
 \right) . 
\end{align}
 Associated with the absolute value of  Bob's measurement outcome $m$, this state is filtered as  
\begin{align}
\rho _{ABE}^{|m|}= \frac{ \hat P_m \rho _{ABE}  \hat P_m } { \tr( \hat P_m \rho _{ABE}  \hat P_m )},
\end{align}
where we defined the projector as \begin{align}
 \hat P_m= \ket{m}\bra{m}_B+\ket{-m}\bra{- m}_B . 
\end{align}

Recall that Alice's bit value $i \in\{0,1\}$ can be determined by projecting her qubit into $\ket{ i}_A$ and Bob's bit value $j \in \{0,1\}$ is determined by projecting his mode into $\ket{(-1)^j |m|}_B$. 
Therefore, Eve's states in Eq.~\eqref{dmA}  can be written as 
\begin{align}
\rho_A^i =  \frac{ \tr_{B}  \bra {i} \rho _{ABE}^{|m|} \ket{i}_A }{ \tr  \bra { i} \rho_{ABE}^{|m|} \ket{ i}_A },  \end{align}
as well as  Eve's states in Eq.~\eqref{dmB}
\begin{align}
\rho_B^j =  \frac{ \tr_{A}  \bra {(-1)^j |m | } \rho _{ABE}^{|m|} \ket{(-1)^j |m |}_B }{ \tr  \bra {(-1)^j |m |} \rho_{ABE}^{|m|} \ket{(-1)^j |m | }_B }.   \end{align}

Since we have traced out the system $D$, it seems difficult to analytically find all eigenvalues of the relevant density operators in the Holevo quantity of Eq.~\eqref{hol} in sharp contrast to the previous approaches \cite{Heid07,Hirano17} where the rank of the density operators was a few, and analytic expressions of all eigenvalues were found. In this paper, we expand the density operator in  a photon-number basis (See  appendix \ref{app-Number-basis}), and determine the Holevo quantity by numerically   finding eigenvalues.

\section{Key rates}
\label{keyrates}

The secure key against collective Gaussian attacks can be calculated from the difference between the information of Alice and Bob in Eq.~\eqref{MIAB} and  Eve's potential knowledge in Eq.~\eqref{hol} \cite{Hirano17}
\begin{align}
\int P(m|\alpha) \max [ I_{AB} - \chi ,0] dm, 
\end{align}
where 
$  P(m| \alpha)$ is given by Eq.~\eqref{ms}, 
and we carry out a postselection process that integrates measurement outcomes  satisfying 
\begin{align}
\max [ I_{AB} - \chi ,0] \ge  0.
\end{align}
   \begin{figure*}[htpb]
  \begin{center}
\includegraphics[width=0.8\linewidth]{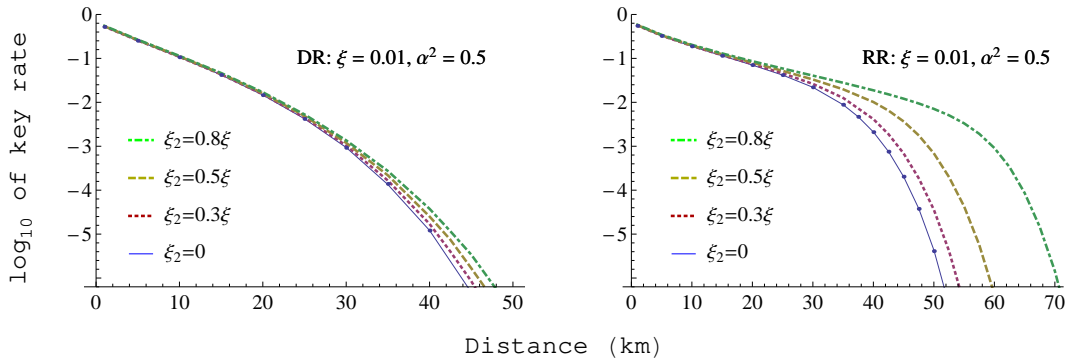}
  \end{center}
  \caption{Key rates for the direct reconciliation (DR) scheme and the reverse-reconciliation (RR) scheme with the photon number $\alpha^2 =0.5 $ and  the total excess noise $\xi=0.01$. The amount of detector's excess noise $\xi_2$ is set to 0, 30\%, 50\% , and 80\%  of the total excess noise $\xi=0.01$. 
The circles on the curve of  $\xi_2 = 0$ are calculated from the analytic result in Ref.~\cite{Hirano17}. }
    \label{figKRPN0.eps}
\end{figure*}

In our previous work \cite{Hirano17}, an analytic expression of the Holevo quantity $\chi$ 
was obtained and a set of the key rates was shown as a function of transmission distance with the total excess noise $\xi= \{ 0.005, 0.01, 0.02 \}$ where  the photon number $|S|^2=\alpha ^2 $ was selected to maximize  the key rate and  the detector  was assumed to be an ideal homodyne detector [In the present notation,  this corresponds to $(\eta , \xi ) =  (\eta_1, \xi_1)$ and $ (\eta_2, \xi_2 ) = (1,0) $]. 
Here we can calculate the key rate for the detector described by a Gaussian channel. We address  the key rate with $\xi_2 >0$ and $\eta_2 =1$ based on the eavesdropping model described in the previous section.

Figure~\ref{figKRPN0.eps} shows a set of numerically calculated key rates with the fixed photon number $ \alpha ^2 =0.5$ and  the total excess noise $\xi=0.01$ for the detector noise $\xi_2 = \{0, 0.3\xi, 0.5\xi , 0.8\xi \}$ and detection efficiency $\eta_2=1$.  Note that the relations of the channel parameters $(\eta_1, \xi_1)$, the detector parameters  $(\eta_2, \xi_2)$, and the total observed transmission   $\eta$  and the total excess noise  $\xi$ are given in Eqs.~\eqref{CPconcate}.  Note also that the value of the photon number $ \alpha ^2 = 0.5$ was close to the optimal photon number that maximizes the key rates in the case of $0.005 \le \xi \le 0.02 $ when  the distance $d$ is located somewhere between $5 $km and $40$km both in the RR and DR schemes \cite{Hirano17}.  

As expected  the decay of the key rate is to some extent delayed as the ratio of the detection noise to the excess noise $\xi_2/\xi $ becomes larger. Notably, this effect is more significant in the case of the RR scheme. For instance, if we compare the solid curve of $\xi_2=0$ and the dot-dash-dot curve of  $\xi_2=0.8\xi $,   we can observe that,  for a given key rate $G > 10^{-6}$, the difference  of distance
  is less   than $5$km for the DR scheme whereas the difference of distance can be more than $15$km for the RR scheme.  
In both case, our result suggests that one can improve the key rate by separately specifying the channel parameters and the detector parameters.
An intuitive reason that the RR scheme has more advantage on specifying the detection noise, is that it becomes harder for Eve to infer Bob's outcome as the detection noise inside Bob's station becomes larger. The fact that the RR scheme maintains its flat decay for a longer distance suggests that  CV-QKD schemes could be incorporated with a long distance network beyond the metropolitan networks.

\begin{figure*}[tbph]
  \begin{center}
\includegraphics[width=\linewidth]{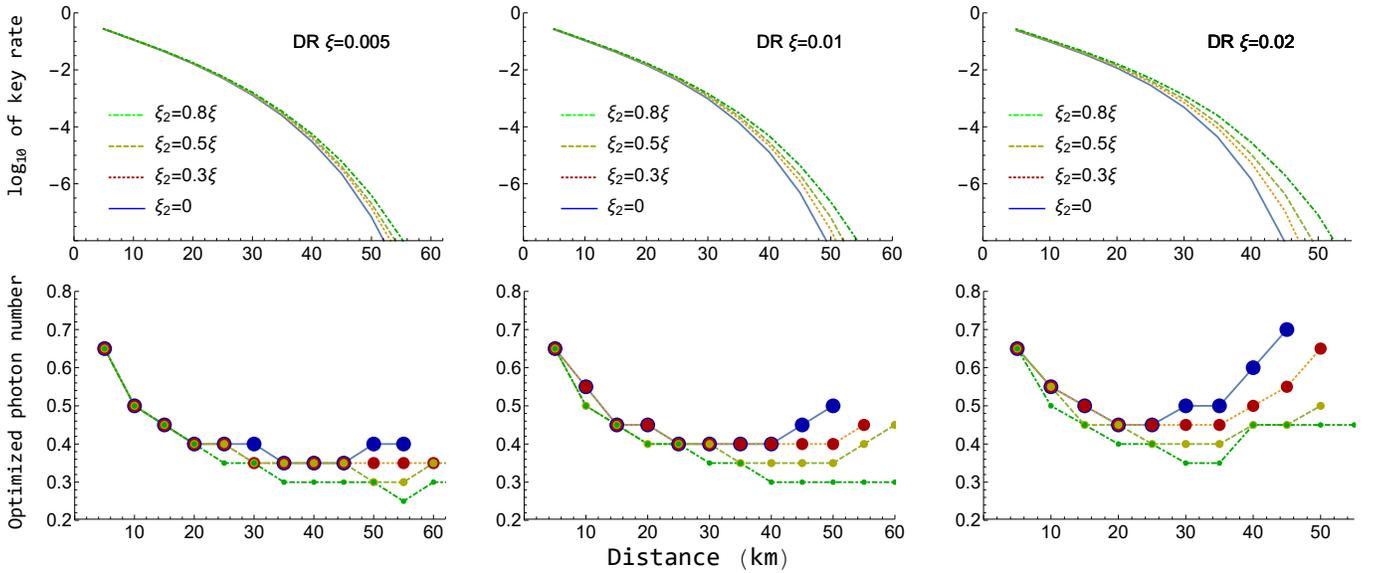}
  \end{center}
  \caption{Key rates for the direct reconciliation (DR) scheme as functions of distance with  the total excess noise $\xi=\{0.005,0.01,0.02\}$.   The photon number $\alpha^2$ is chosen to maximize the key rate with $0.05$ steps.  The amount of  detector's excess noise $\xi_2$ is set to 0, 30\%, 50\% , and 80\%  of the total excess noise $\xi$.}
    \label{figDRGraph.eps}
\end{figure*}

Since initially there is no specific constraint on the average photon number $\alpha^2 $ of  four coherent states, we can choose this parameter so as to extract higher key rate.  The upper panels of Fig.~\ref{figDRGraph.eps} show  the maximum key rates of the DR scheme for  the excess noise $\xi =\{0.005,0.01,0.02 \}$ when the photon number $  \alpha  ^2 $ is optimized.  The key rate is basically calculated in each $5$km-distance step  and photon number step of  $0.05$  from $d=5$km. The lower panels of Fig.~\ref{figDRGraph.eps} show  the corresponding optimized values of the photon number $\alpha^2$.

\begin{figure*}[tbph]
  \begin{center}
\includegraphics[width=\linewidth]{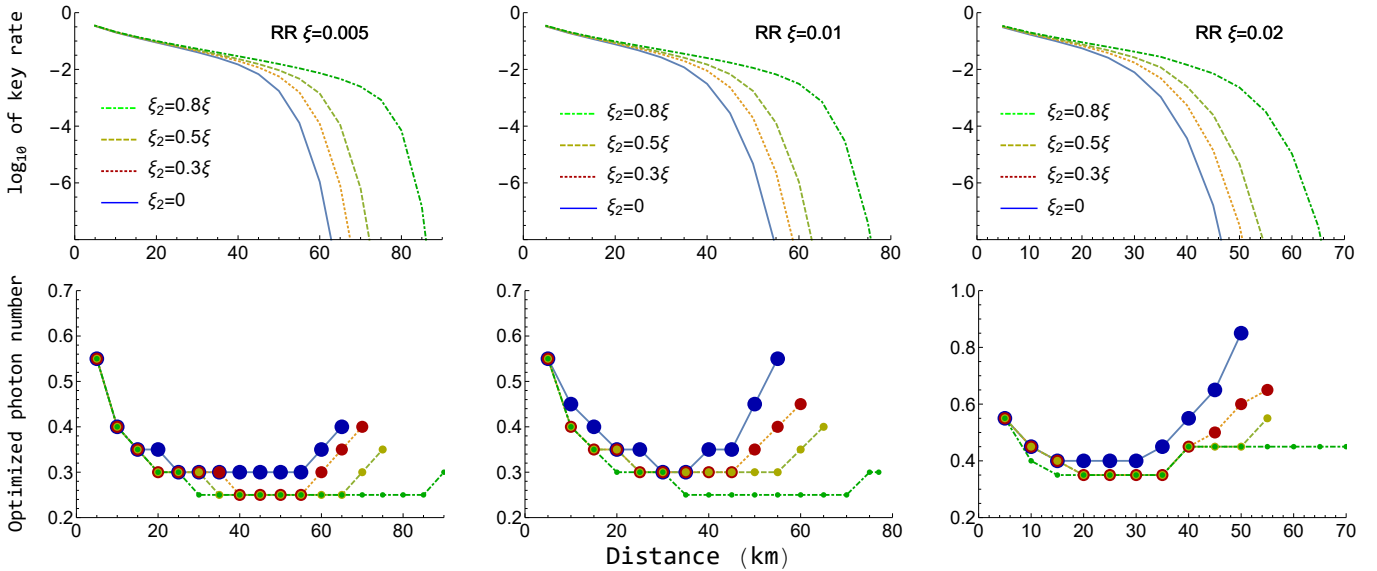}
  \end{center}
  \caption{Key rate for the reverse-reconciliation (RR)  scheme as functions of distance with  the total excess noise $\xi=\{0.005,0.01,0.02\}$.   The photon number $\alpha^2$ is chosen to maximize the key rate with $0.05$ steps.  The amount of  detector's excess noise $\xi_2$ is set to 0, 30\%, 50\% , and 80\%  of the total excess noise $\xi$.  }
    \label{figRRGraph.eps}
\end{figure*}

The upper panels of Fig.~\ref{figRRGraph.eps} show  the maximum key rates of the RR scheme for  the excess noise $\xi =\{0.005,0.01,0.02 \}$. 
The key rate is basically calculated in each $5$km-distance steps  and photon number steps of  $0.05$  from $d=5$km. The lower panels of Fig.~\ref{figRRGraph.eps} show  the corresponding optimized values of the photon number. For a fixed total excess noise $\xi$, one can see that the shallow decay of the key rates sustains longer as the potion of the detection noise becomes larger. Thereby, depending on the distance we can obtain much better key rate by carefully calibrating the detection noise and the channel noise, separately.

In optimizing the photon number $\alpha^2$, it  could be helpful  to consider that there are three typical regions on the distance as follows: (i) For very shorter distances with smaller loss, Eve's information could be small. We thus can use a relatively higher value of the photon number for  such as $d \sim 10$km. (ii) For middle distances, the optimal photon number becomes smaller and its value is almost flat as a function of distance. The flatness suggests that the key rates are insensitive to the difference of the photon number with regards to the scale of such as 0.5 photon number. This could be favorable for a practical implementation when a fine control of the photon number is difficult. (iii) Finally for toward long  distances, an increase of the optimal photon number is observed. In this regime, the key rate would drop more rapidly when one deviates the value of the photon number from the optimal value. Thereby, our numerical search of the optimal point could take a relatively long time, and the key rate itself is typically very low.

As a summary, both in the DR and RR schemes, the typical behavior of the key rates with optimized photon number  is no much different from the behavior of the key rate with the case of the fixed photon number $\alpha ^2 =0.5 $ in Fig.~\ref{figKRPN0.eps} for  all cases of  the excess noise $\xi =\{0.005,0.01,0.02 \}$.  As far as our model described, the critical point in experiments is to suppress the excess noise both in the transmission process and the detection process as low as possible.   
An essential observation is that, for given total excess noise $\xi$, the optimal photon number becomes smaller as the portion of  detection noise becomes larger. There is a trade-off relation that  a smaller photon number implies lower information for Alice and Bob, and larger information leakage for Eve. Hence, one could not simply suggest that a better choice of the value of the initial photon number is high or low when  the parameters change. In the present case of  the numerically simulated regime, 
it turn out that a smaller photon number such as $\alpha ^2 \in (0.3,0.5)$ was widely feasible.  In this respect, systematic numerical study would be more important to find out  efficient settings in operating our QKD schemes.

\section{Conclusion} 
\label{Conclusion}
We have developed a method to calculate the key rate of a CV-QKD scheme using four coherent states for a model of Gaussian attacks. We assume that the transmission channel and detection process are described by a pair of symmetric Gaussian channels. The key rate against collective Gaussian attacks can be essentially calculated for any set of the channel parameters.  
We use two entangling cloners to describe the two of the noise sources. One entangling cloner is assumed to be controlled by Eve as  usual and represents the information leakage for her. The other entangling cloner represents the detection noise and is  located inside Bob's station. Thereby, the signal outgoing to  detector's environment is simply traced out and gives no information leakage to Eve.  We showed numerically determined key rates as functions of distance for a couple of possible combinations of excess noise parameters. By separating the channel noise and detection noise,  
we can always extract larger key compared with the case that all loss and noises are induced by Eve's interference. The improvement of the key rates turns out to be more significant for the RR schemes where the increase of inaccessible noises at the detector is thought to make Eve more difficult to infer Bob's outcomes.  Note that our study is for an asymptotic key rate and  limited to the case of eavesdropping models that induce symmetric Gaussian noise and linear loss. Since there is no proof that the Gaussian attack is the optimal attack for CV-QKD schemes with discrete modulation, achievable key rates in general eavesdropping scenario could be significant lower than the present key rates \cite{Zhao09}.  
It may be worth noting that our method developed here would be applicable to the protocols proposed in Ref.~\cite{Namiki2006}, which include an efficient four-state protocol based on diagonally modulated coherent states, i.e., the coherent states in the form of  $\ket{\pm \alpha \pm i \alpha }$.

\begin{acknowledgments}
This work was supported by ImPACT Program of Council for Science, Technology and Innovation (Cabinet Office, Government of Japan),  JSPS KAKENHI
Research (B) (Grants No. JP18H01157), and  JSPS KAKENHI
Research (S) (Grants No. JP18H05237).  
\end{acknowledgments}

\appendix
\section{Number-basis representation of $\rho_{S,m} $ in Eq.~\eqref{RhoSm}} \label{app-Number-basis}
Given a matrix representation of the density operator  $\rho_{S,m} $ in Eq.~\eqref{RhoSm}, we can determine its eigenvalues numerically. From the  set of eigenvalues we can determine the Von-Neumann entropy and thus the Holevo quantity  $\chi$ in  Eq.~\eqref{hol}. In this appendix we present a basic calculation to expand the relevant density operator in the number basis so as to determine the matrix elements.  

From the wave function of Eq.~\eqref{WF1} we can carry out the integration of $Y$ in $\rho_{S,m} $ of Eq.~\eqref{RhoSm}. For instance, we can write  
\begin{align}
 & \bra{x_1,x_2}\rho_{S,m} \ket{x_3,x_4 } e^{-\sum_{i=1}^4x_i^2}  \nonumber \\
=& \sqrt{\frac{4}{\pi} \frac{1}{V_2+\frac{1}{V_2}} } \int dY  \varphi_Y (x_1,x_2)  \varphi_Y (x_3,x_4) 
e^{ -\frac{4 Y^2}{V_2+\frac{1}{V_2}} -\sum_{i=1}^4x_i^2}  \nonumber \\
= &  \sqrt{\frac{4}{\pi} \frac{1}{V_2+\frac{1}{V_2}} } \sqrt{\frac{8}{\pi ^2 \mathcal{P}\eta _2 (1-\eta _2 )} } 
  e^{- \vector{ x} ^t A \vector{ x}  + 2 \vector{ x} ^t \vector{ b} +c }, \label{nono1}
\end{align}
where  
\begin{align}
\mathcal{P}=\left( V_1 +\frac{1}{V_1 } \right) \frac{1-\eta _1 }{\eta _2 } 
+\frac{2\eta _1 }{\eta _2 } 
+\frac{4}{V_2 +\frac{1}{V_2 } } \frac{1}{1-\eta _2 },  
\end{align}and we defined $(A, \vector{b},c)$ so as to get the quadratic form of
 $  \vector{ x} =(x_1,x_2,x_3,x_4)^t $. 
The elements of  $(A, \vector{b},c)$ are determined  as 
\begin{align}
A=
\left( 
\begin{array}{cccc}
\alpha _1 & \alpha _3 & \alpha _4 & \alpha _6 \\
\alpha _3 & \alpha _2 & \alpha _6 & \alpha _5 \\
\alpha _4 & \alpha _6 & \alpha _1 & \alpha _3 \\
\alpha _6 & \alpha _5 & \alpha _3 & \alpha _2 
\end{array}
\right) , \quad 
\vector{b}
 =\left( 
\begin{array}{c}
\beta _1 \\
\beta _2 \\
\beta _1 \\
\beta _2 
\end{array}
\right) , \\
c = -\frac{4}{V_2 +\frac{1}{V_2 } } \frac{m ^{ 2} }{1-\eta _2 }+\frac{r^2 }{\mathcal{P}} 
-2S^2 ,
\end{align}
with the help of the coefficients
\begin{eqnarray}
\alpha _1 &=&1-\frac{p^2 }{\mathcal{P}} +\frac{1}{2} \left(V_1 +\frac{1}{V_1 } \right) , \\
\alpha _2 &=&1-\frac{q^2 }{\mathcal{P}} +(1-\eta _1 )+\frac{\eta _1 }{2} \left( V_1 +\frac{1}{V_1 } \right) , \\
\alpha _3 &=&-\frac{pq}{\mathcal{P}} +\frac{\sqrt{\eta _1 }}{2} \left( V_1 +\frac{1}{V_1 } \right) , \\
\alpha _4 &=&-\frac{p^2 }{\mathcal{P}} , \quad  
\alpha _5  = -\frac{q^2 }{\mathcal{P}} ,  \quad 
\alpha _6  = -\frac{pq}{\mathcal{P}} , \\
\beta _1 &=&\frac{pr}{\mathcal{P}} ,  \quad  
\beta _2  = \frac{qr}{\mathcal{P}} +\sqrt{1-\eta _1 }S, 
\end{eqnarray}
and the constants independent of the elements of $\vector{ x} $ 
\begin{eqnarray}
p&=&-\frac{1}{2} \left( V_1 -\frac{1}{V_1 } \right) \sqrt{\frac{1-\eta _1 }{\eta _2 } } , \\
q&=&\left[ 1-\frac{1}{2} \left( V_1 +\frac{1}{V_1 } \right) \right] 
\sqrt{\frac{\eta _1 (1-\eta _1 )}{\eta _2 } } , \\
r&=&-2\sqrt{\frac{\eta _1 }{\eta _2 } } S-\frac{4}{V_2 +\frac{1}{V_2 } } \frac{m}{1-\eta _2 }.  
\end{eqnarray}
Note that  $(V_1, V_2)$ are defined in Eqs.~\eqref{V1}~and~\eqref{V2}. They are equivalently written as 
\begin{align}
V_i&= w_i + \sqrt{w_i^2-1}, \nonumber \\ 
w_i&=\frac{1-\eta_i+\xi_i}{1-\eta_i} \label{vw2},
\end{align} for $  i =1,2 $.

In order to reach the number state representation, 
let us write the  wave functions of number states  as  
\begin{align}
\ket{n} =& \int_{-\infty }^\infty  u_n (x ) \ket{x} d x , \nonumber \\ 
u_n(x ) := & \left( \frac{2}{\pi} \right) ^{1/4} \frac{1}{\sqrt{2 ^{n } n !}} 
 H_n(\sqrt 2 x )  e^{ - x^2},\nonumber \\
H_n(x):=& (-1)^n e^{x^2} \frac{d^n e ^{-x^2 }}{dx^n } .
\end{align}
With these expressions and Eq.~\eqref{nono1} we can obtain the matrix element of $\rho_{S,m}$ as
\begin{align}
& \bra{n_1,n_2}\rho_{S,m} \ket{n_3,n_4 }\nonumber \\ 
 =&\frac{2}{\pi} \iiiint \textrm{d}^4 x\prod_{i=1}^4  \frac{H_{n_i}(\sqrt{2}x_i) e^{-x_i^2}}{\sqrt{2 ^{n_i} n_i!}}  \bra{x_1,x_2}\rho_{S,m} \ket{x_3,x_4 } \nonumber \\ 
= &\frac{2}{\pi} \sqrt{\frac{4}{\pi} \frac{1}{V_2+\frac{1}{V_2}} } \sqrt{\frac{8}{\pi ^2 \mathcal{P}\eta _2 (1-\eta _2 )} }   \nonumber \\ 
&\times  \iiiint_{-\infty}^{\infty} \prod_{i=1}^4  \frac{dx_i {H_{n_i}(\sqrt{2}x_i) }}{\sqrt{2 ^{n_i} n_i!}}    e^{- \vector{ x} ^t A \vector{ x} + 2 \vector{ x} ^t \vector{ b} +c }. 
\end{align}

Now, we can numerically determine the matrix elements. In our numerical approach, the matrix $A$ is diagonalized. 
Then, we have a simpler quadratic form as   
\begin{align}
&\vector{ x} ^t A \vector{ x}  - 2 \vector{ x} ^t \vector{ b} - c  \nonumber \\ 
=& (A^{\frac{1}{2}} \vector{ x}  - A^{- \frac{1}{2}}\vector{ b}  )^t (A^{\frac{1}{2} }\vector{ x}  - A^{- \frac{1}{2}}\vector{ b}  ) -\vector{b} ^t A^{-1}\vector{ b} -c \nonumber \\
=& \vector{y}^t\vector{y}     - c^\prime   ,
\end{align}
where 
\begin{align}
  \vector{ y } =&(y_1,y_2,y_3,y_4)^t 
= A^{\frac{1}{2}} \vector{ x}  + A^{- \frac{1}{2}}\vector{ b},   \label{xdir} \\    c^\prime= & c +\vector{b} ^t A^{-1}\vector{ b} . 
\end{align}
The volume element  associated with this change of variables is given by $ \textrm{d}^4x= \left|\frac{ \partial x_i}{\partial y_j} \right|  \textrm{d}^4 y  =   \det( A^{-1/2})   \textrm{d}^4 y   $. 
From the property of the determinant $\det ( a) \det (b ) = \det (ab)$, 
we can write $\det(A^{1/2}) = \sqrt{\det(A)} $. 
Using these relations 
we have \begin{align}
& \bra{n_1,n_2}\rho_{S,m} \ket{n_3,n_4 }\nonumber \\ 
= &\frac{2}{\pi} \sqrt{\frac{4}{\pi} \frac{1}{V_2+\frac{1}{V_2}} } \sqrt{\frac{8}{\pi ^2 \mathcal{P}\eta _2 (1-\eta _2 )} } \nonumber \\ 
&\times  \frac{e^{- c^\prime}  }{\sqrt {\det (A)}}  \iiiint_{-\infty}^{\infty}  \prod_{i=1}^4  \frac{dy_i {H_{n_i}(\sqrt{2}x_i) }}{\sqrt{2 ^{n_i} n_i!}}    e^{-  y_i^2 } ,  \end{align}
where $ \{x_i \} $ will be represented in terms of $\{y_i\}$ through the inverse of Eq.~\eqref{xdir}, namely,  \begin{align}
\vector{ x } =&A^{- \frac{1}{2}} \vector{ y}  - A^{- 1} \vector{ b}.    \end{align}
Since the product $ \prod_i {H_{n_i}(\sqrt{2}x_i) }$ is a polynomial of $(y_1,y_2,y_3,y_4)$  
 we can carry out the integration 
by recursively using the well-known formula for the Gaussian integral: 
\begin{align}
\int_{-\infty}^{\infty} y_i^{2n} e^{-y_i ^2} dy_i =&  \sqrt \pi  \frac{(2n-1)!!  }{2^n   }, \nonumber \\ 
\int_{-\infty}^{\infty} y_i^{2n+1 } e^{-y_i ^2} dy_i =&  0 ,
\end{align} for $n =0,1,2, \cdots$.

\end{document}